 \documentclass[final,5p,times,twocolumn]{elsarticle}

\usepackage{makecell}
\usepackage{latexsym}
\usepackage{amssymb, amsmath, bm, physics}
\usepackage{subcaption}
\usepackage{mathptmx}
\usepackage{flushend}
\usepackage{booktabs}
\usepackage{lipsum}
\usepackage{adjustbox}
\usepackage{tabularx}
\usepackage[colorlinks,citecolor=blue,urlcolor=blue,linkcolor=blue]{hyperref}
\usepackage{orcidlink}
\usepackage{multicol}

\journal{Nuclear Physics B}

\begin{document}

\begin{frontmatter}

\title{Constraining fractionality using some observational tests}

\author[1]{H. Moradpour \orcidlink{0000-0003-0941-8422}}
\affiliation[1]{organization={Research Institute for Astronomy and Astrophysics of Maragha (RIAAM)},%Department and Organization
            addressline={University of Maragheh}, 
            city={Maragheh},
            postcode={55136-553}, 
          %  state={PE},
            country={Iran}}            
\ead[1]{h.moradpour@riaam.ac.ir}
\author[2,3,4]{S. Jalalzadeh \orcidlink{0000-0003-4854-2960}} 
\affiliation[2]{organization={Izmir Institute of Technology},%Department and Organization
            addressline={Department of Physics}, 
            city={Urla},
            postcode={35430}, 
            state={Izmir},
            country={Türkiye}}
\affiliation[3]{organization={Dogus University},%Department and Organization
            addressline={Department of Physics}, 
            city={Dudullu-Ümraniye},
            postcode={34775}, 
            state={Istanbul},
            country={Türkiye}}
\affiliation[4]{organization={Khazar University},%Department and Organization
            addressline={Center for Theoretical PhysicsCenter for Theoretical Physics}, 
            city={41 Mehseti Street},
            postcode={AZ1096}, 
            state={Baku},
            country={Azerbaijan}}               
\ead[2]{shahramjalalzadeh@iyte.edu.tr}      

\author[1]{R. Jalalzadeh \orcidlink{0000-0002-6110-3981}}         
\ead[3]{r.jalalzadeh@riaam.ac.ir}

\author[1]{A. H. Ziaie \orcidlink{0000-0001-7029-134X}}    
\ead[3]{ah.ziaie@riaam.ac.ir}

%%%%%%%%%%%%%%%%%%%%%%%%%%%%%%%%%%%%%%%%%%%%%%%%%%%%%%%%%%%%%%%%%%%%%%%%%%%%%%%%%%%%%%%%%%%%%%%%%%%%%%%%%%%%%%%%%
\begin{abstract}
 Recently, a fractional version of the Schwarzschild-Tangherlini black hole with a fractal horizon has been introduced. Motivated by the key role of the Schwarzschild solution in gravitational and
astrophysical studies, some consequences of this fractional-fractal generalization of the Schwarzschild black hole have been investigated. In this line, the corresponding i) Shapiro and Sagnac time delays, ii) shadow, iii) orbital precession, and iv) gravitational lensing are studied and confronted with observational data. MCMC analysis also unveils i) the potential of this metric in dealing with the Solar-system tests and ii) the necessity of studying fractional spacetimes and objects.
\end{abstract}



%\begin{keyword}

%\end{keyword}

\end{frontmatter}

%\tableofcontents


\section{Introduction}

Mathematically, \textit{fractional calculus} equips physicists with a useful tool for studying fractal systems, and indeed, has diverse applications in physics \cite{Herrmann:2011zza, Laskin:1999tf, Laskin:2002zz, Tarasov:2005orz}. Fractal structures are predicted and expected in various gravitational, astrophysical, and cosmological systems \cite{SylosLabini:1997jg, Capozziello:2009qu, Verevkin:2011qe, Roy:2009tr, Asch, allen, Majumder:2017rsy} and expectedly, by employing fractional calculus, various attempts have been started to study such systems \cite{deOliveiraCosta:2023srx, Junior:2023fwb, Jalalzadeh:2023mzw, Socorro:2023xmx, Gonzalez:2023who, Garcia-Aspeitia:2022uxz, Calcagni:2019ngc, Calcagni:2017via, Calcagni:2020tvw, Giusti:2020rul, Mureika:2006tz, Shchigolev:2010vh, Calcagni:2009kc, Varieschi:2021rzk}. Recently, motivated by the pivotal role of the Schwarzschild black hole in our understanding of various phenomena, a fractional Schwarzschild-Tangherlini solution has been introduced whose horizon has a fractal structure \cite{Jalalzadeh:2025uuv}.

The fractional static and spherically symmetric Schwarzschild--Tangherlini line element is reported in the form of \cite{Jalalzadeh:2025uuv}

\begin{equation}\label{h2}
ds^2=-f(r)c^2dt^2+\frac{dr^2}{f(r)}+r^2d\Omega^2_{D-2}=-d\tau^2,
\end{equation}

\noindent in which the angular sector is also described by a fractional (Hausdorff) horizon geometry

\begin{multline}\label{Fline}
d\Omega^2_{D-2}=\sin^2\theta
d\varphi^2+\Big\{\cos^2\theta+\\\left(\frac{\pi^\frac{D-3}{2}}{\Gamma(\frac{D-3}{2})}\right)^2|\cos\theta|^{2(D-4)}\sin^2\theta
\Big\}d\theta^2,
\end{multline}

\noindent where

\begin{equation}
\begin{split}\label{h4}
&f(r)=1-\frac{\gamma}{r^{D-3}},\qquad
\gamma\equiv \frac{16\pi \tilde G M}{(D-2)~c^2},\\
&\Omega_{D-2}=\frac{2\pi^{\frac{D-1}{2}}}{\Gamma(\frac{D-1}{2})},\qquad\tilde{G}=\mathcal{A}~l_p^{D-4}G,\\
&\mathcal{A}=\frac{(D-2)^{D-2}~B\big(\frac{1}{2},\frac{2D-5}{2(D-3)}\big)^{D-3}}{8\pi^{D-2}},
\end{split}
\end{equation}

\noindent in which $l_p$ denotes the Planck length and in the unit of $c=\hbar=1$, we have $l_p^2=G$. Also, $D$ $(3<D\leq4)$ represents the fractional dimension related to the Levy's fractional parameter ($\alpha$) as $D=\alpha/2+3$, and obviously, the dimension of $\gamma$ is equal to $L^{D-3}$. Moreover, $G$ ($\tilde G$) is the gravitational constant in $D=4$ (effective gravitational constant for $D\neq4$), while $\Gamma(\frac{D-3}{2})$ and $B(a,b)$ denote the Gamma function and the Beta function

\begin{equation}
B(z_1,z_2)=2\int_0^\frac{\pi}{2}\sin^{2z_1-1}s~\cos^{2z_2-1}s~ds,
\end{equation}

\noindent respectively. From now on, to keep things simple, we will work with natural units, unless we need to revert to SI units. Consequently, a $D$-dimensional spacetime supported by a fractional Schwarzschild--Tangherlini black hole whose surface has a fractal structure with dimension $D-2$ located at $r_{h}=(\frac{16\pi\tilde{G}M}{(D-2)})^{\frac{1}{D-3}}$ is dealt with \cite{Jalalzadeh:2025uuv}. The Schwarzschild spacetime and the corresponding horizon radius are easily recovered for $D=4$ ($\alpha=2$).

Undoubtedly, the observational methods for detecting and distinguishing such a solution from a non-fractal one are of importance, as their existence can dramatically affect our understanding of the Universe and spacetime. Time delays measurements are from these methods. Distinct objects bend spacetime differently that may affect observations. Correspondingly, the travel time of a light beam in a gravitational field is longer than its travel time in a flat space-time. The latter inspires a gravitational time delay called the Shapiro time delay \cite{Shapiro:1964uw, Shapiro:1968zza, Shapiro:1971iv,cl1}. It has been shown that this time delay can be used to test the gravitational theories, higher dimensions, distinguish a naked singularity from a black hole, etc \cite{Shapiro:1964uw, Shapiro:1968zza, Shapiro:1971iv, cl1, cl2, Azar:2023fsv}.

There is also another time delay called the Sagnac time delay \cite{Sagnac:1899, Sagnac:1913} that, despite its difficulties \cite{Azar:2023fsv}, seems useful to study singularities, the extra dimensions, and detecting dark matter in the Solar system \cite{Ashtekar:1975wt, Anandan:1981zg, Rizzi:2003dh, Ruggiero:2004iaq, Benedetto:2019ltm, Ziaie:2022zmz, Souza:2024ltj}. Finally, due to the amazing gravity of black holes, they can noticeably cast their shadow on the background \cite{Chandrasekhar:1985kt}, and expectedly, efforts to study the black hole shadows are increasing following the recent observational confirmation of the shadow \cite{EventHorizonTelescope:2019ggy, EventHorizonTelescope:2019dse, EventHorizonTelescope:2022wkp, Vagnozzi:2022moj, Gralla:2019xty, Perlick:2021aok, Zahid:2025cfu}. Gravitational lensing, as well as the orbit precession, are other ways to confirm a gravitational theory as well as the existence of a geometry \cite{dInverno:1992gxs, Weinberg:1972kfs}.

Accordingly, the main concern of this paper is to pave the way for tracing the footprints of fractionality by studying the Shapiro and Sagnac time delays as well as the shadow, gravitational lensing, and orbit precession corresponding to metric~(\ref{h2}). In this line, the Shapiro and Sagnac time delays are addressed in the next section. The bending of null and time-like geodesics, which lead us to the black hole shadow as well as the gravitational lensing, and the precession of planetary orbits, respectively, are investigated in the third section. A summary is also presented in the last section.

%%%%%%%%%%%%%%%%%%%%%%%%%%%%%%%%%%%%%%%%%%%%%%%%%%%%%%%%%%%%%%%%
\section{Gravitational time delays}

In the presence of gravity, the spacetime is curved. Therefore, a photon traveling between two points in a curved spacetime takes more time compared to a similar journey in the flat spacetime. It is the backbone of time delays in gravitational fields.

\subsection{Sagnac time delay}

If a sender/receiver rotates around an object, a time difference
between the received signals is seen depending on their movement
direction. This is the Sagnac time delay that seems powerful to
distinguish naked singularities from non-naked one, and survey the higher dimensions and related phenomena \cite{Sagnac:1899, Sagnac:1913, Ashtekar:1975wt, Anandan:1981zg, Rizzi:2003dh, Ruggiero:2004iaq, Benedetto:2019ltm, Ziaie:2022zmz}. To verify the ability of this effect to detect the fractional dimension, we focus on
metric~(\ref{h2}) and compare the result with that of the ordinary Schwarzschild black hole. Without loss of generality, we confine ourselves to the plane $\theta=\frac{\pi}{2}$. Therefore, we deal with a Sagnac sender/receiver apparatus with angular velocity $\omega_0$ and $d\theta=0$, rotating around the black hole.
Now, applying $(t,\phi)\to (t,\phi+\omega_0 t)$ on the equatorial plane, the nonzero metric components become

\begin{equation}
g_{00}= -\big[f(r)-r^2\omega_0^2\big],\quad
g_{0\phi}= r^2 \omega_0,\quad
g_{\phi\phi}= r^2.
\end{equation}

\noindent Hence, $A_\phi \equiv g_{0\phi}/|g_{00}| = \dfrac{r^2\omega_0}{f(r)-r^2\omega_0^2}$,

\begin{equation}
\Delta t = 2\sqrt{|g_{00}|}\int_0^{2\pi}\!A_\phi\,d\phi
= 2\pi\,\frac{2 r^2\omega_0}{\sqrt{f(r)-r^2\omega_0^2}},
\end{equation}

\noindent and consequently, for a circular path at $r=R$ with $A\equiv 4\pi R^2$ and $f(R)=1-\gamma/R^{D-3}$, the Sagnac time factor is achieved as

\begin{equation}\label{S04}
t_F = A\,\omega_0\left(1-\frac{\gamma}{R^{D-3}}-R^2\omega_0^{\,2}\right)^{-1/2}.
\end{equation}

\noindent Expectedly, the Sagnac time delay of a Schwarzschild black hole ($t_{Sch}$) \cite{Tartaglia:1998rh, Sakurai:1980te, Anandan:1981zg,Rizzi:2003dh, Ruggiero:2004iaq, Ruggiero:2005nd, Ziaie:2022zmz}

\begin{eqnarray}\label{S4}
t_{Sch}=A\omega_0\left(1-\frac{2GM}{R}-R^2\omega_0^2\right)^{-\frac{1}{2}},
\end{eqnarray}

\noindent is also recovered by inserting $D=4$ (the desired limit) in Eq.~(\ref{S04}). Therefore, the surface properties of an object $M$ are verifiable
by following the trace of $\frac{\gamma}{R^{D-3}}$ in justifying observations. To obtain an estimate of the power of the Sagnac test in detecting the value of $D$, it is useful to note that for $3<D<4$, a numerical analysis shows that

\begin{eqnarray}
\delta(D)\equiv\frac{\big|t_{Sch}-t_F\big|}{t_{Sch}}\approx\mathcal{O}(10^{-2}).
\end{eqnarray}

\noindent Therefore, given the accuracy of atomic time measurements \cite{Souza:2024ltj}, it seems likely that, in addition to looking for dark matter in the Solar system \cite{Souza:2024ltj}, future Sagnac apparatus will also be able to measure the value of $D$.

%at least in large gravitational fields, where deviations from general relativity and its common solutions such as the Schwarzschild spacetime become important.

%%%%%%%%%%%%%%%%%%%%%%%%%%%%%%%%%%%%%%%%%%%%%%%%%%%%%%%%%%%%%%%

\subsection{Shapiro time delay}

\begin{figure}[h]
\centering
\includegraphics[width=0.8\linewidth]{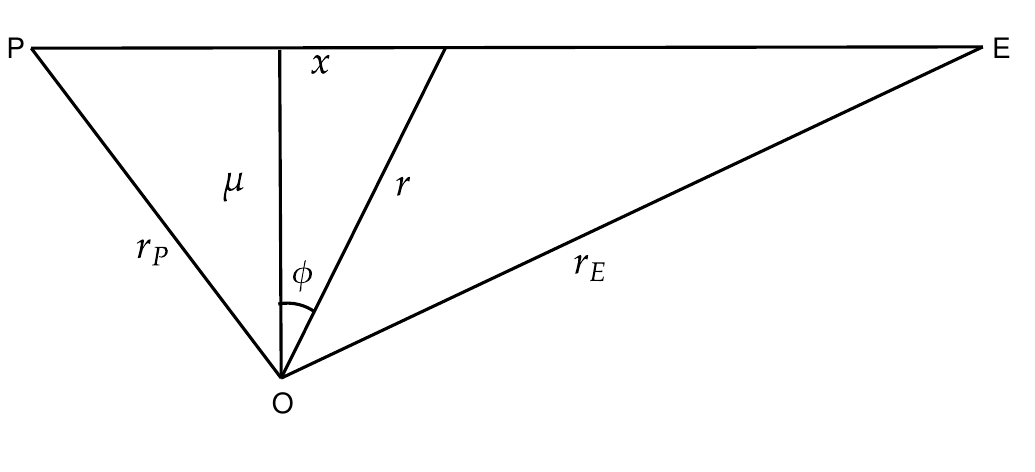}
\caption{The photon path (from $P$ to $E$).} \label{figh1}
\end{figure}

Correspondingly, for a photon following the null geodesics of~(\ref{h2}), there is a time delay in the travel time from the point $P$ towards $E$ compared to the one that does not feel the object $O$ during its journey towards $E$ from $P$ (a flat spacetime). In Fig.~\ref{figh1}, the corresponding setup has been provided in which $\mu$ is the vertical distance of the photon's path from the object $O$. This time delay is called the Shapiro time delay \cite{Shapiro:1964uw, Shapiro:1968zza, Shapiro:1971iv, cl1, cl2}, which is useful in distinguishing black holes and naked singularities from each other, examining the dimensions of spacetime, and objects, as well as studying motions in higher dimensions, if they exist \cite{Azar:2023fsv}.

Therefore, following
Ref.~\cite{Azar:2023fsv}, and by considering the equatorial plane ($\theta=\frac{\pi}{2}$) where $g_{\phi\phi}=r^2$, a null ray ($ds=0$) satisfies

\begin{equation}\label{4}
0= -f(r)\,dt^2 + f(r)^{-1}dr^2 + r^2 d\phi^2,
\end{equation}

\noindent and thus, the Shapiro time ($dt$) of photon travel is calculated as

\begin{equation}\label{h5}
\left(\frac{dt}{dr}\right)^2=f(r)^{-2}\left\{1+\frac{\mu^2}{r^2-\mu^2}f(r)\right\},
\end{equation}

\noindent where $d\phi=\frac{\mu}{r\sqrt{r^2-\mu^2}}dr$, obtained from Fig.~(\ref{figh1}), has been employed. Expanding to $\mathcal O(\mu^2)$,

\begin{multline}\label{h6}
\left(\frac{dt}{dr}\right)^2\simeq\\\frac{r^2}{r^2-\mu^2}
\left\{1+\frac{2\gamma}{r^{D-3}}-\mu^2\left(\frac{\gamma}{r^{D-1}}+\frac{2\gamma^2}{r^{2(D-2)}}\right)\right\},
\end{multline}

\noindent and thus

\begin{multline}\label{h7}
dt\simeq\\\frac{rdr}{\sqrt{r^2-\mu^2}}\left\{1+\frac{\gamma}{r^{D-3}}-\mu^2(\frac{\gamma}{2r^{D-1}}+\frac{\gamma^2}{r^{2(D-2)}})\right\}.
\end{multline}

\noindent and the first (flat) term yields $\sqrt{r_E^2-\mu^2}+\sqrt{r_P^2-\mu^2}$, nothing but the Newtonian result. Moreover, since $r=\sqrt{x^2+\mu^2}$, one gets

\begin{equation}
\int \frac{r\,dr}{\sqrt{r^2-\mu^2}}\frac{1}{r^{D-3}}
= \mu^{3-D}\, {}_2F_1\!\left(\frac12,\frac{D-3}{2};\frac{3}{2}; -\frac{x^2}{\mu^2}\right),
\end{equation}

\noindent where ${}_2F_1$ is the Gauss hypergeometric function. Since the difference with the Schwarzschild metric becomes noticeable from the second term ($\frac{\gamma}{r^{D-3}}$), from now on, we shall only focus on the first two terms of Eq.~(\ref{h7}). Bearing $r=\sqrt{x^2+\mu^2}$ and $D=4$ in mind, the Schwarzschild result \cite{dInverno:1992gxs}

\begin{multline}\label{h8}
T_{Sch}=\left\{\sqrt{r_E^2-\mu^2}+\sqrt{r_P^2-\mu^2}\right\}+\\ 2GM\ln\left\{(\sqrt{r_E^2-\mu^2}+r_E)(\sqrt{r_P^2-\mu^2}+r_P)/\mu^2\right\}.
\end{multline}

\noindent is recovered. Therefore, for calculating the Shapiro time delay, one should subtract the Newtonian time travel from $T_{Sch}$ to reach

\begin{eqnarray}\label{h81}
\delta T_{Sch}&=&2GM\ln[(\frac{\sqrt{r_E^2-\mu^2}+r_E)(\sqrt{r_P^2-\mu^2}+r_P)}{\mu^2}]\nonumber\\&\simeq&2GM\ln(\frac{4r_Er_P}{\mu^2}),
\end{eqnarray}

\noindent in which the last expression is supported by considering the $r_E\gg\mu$ and $r_P\gg\mu$ assumptions in agreement with the weak field limit. For other values of $D$, we have

\begin{multline}
\label{h9}
T_F=\left\{\sqrt{r_E^2-\mu^2}+\sqrt{r_P^2-\mu^2}\right\}+\\
\frac{\gamma}{\mu^{D-3}}\Big\{\sqrt{r_E^2-\mu^2}\ _{2}F_1(\frac{1}{2},\frac{D-3}{2};\frac{3}{2};-\frac{r_E^2-\mu^2}{\mu^2})+\\
\sqrt{r_P^2-\mu^2}\ _{2}F_1(\frac{1}{2},\frac{D-3}{2};\frac{3}{2};-\frac{r_P^2-\mu^2}{\mu^2})\Big\},
\end{multline}

\noindent that recovers Eq.~(\ref{h8}) for $D=4$ and finally leads to

\begin{eqnarray}\label{h91}
\delta T_F&\simeq&\frac{\gamma}{4-D}\big(r_E^{4-D}+r_P^{4-D}\big)\\&+&\frac{\gamma\mu^2(D-3)}{2(D-2)}\big[(\frac{1}{r_E})^{D-2}+(\frac{1}{r_P})^{D-2}\big],\nonumber
\end{eqnarray}

%\begin{eqnarray}\label{h91}
%\delta T_F&=&\frac{\gamma}{\mu^{D-3}}\Big\{\sqrt{r_E^2-\mu^2}\ _{2}F_1(\frac{1}{2},\frac{D-3}{2};\frac{3}{2};-\frac{r_E^2-\mu^2}{\mu^2})\nonumber\\&+&
%\sqrt{r_P^2-\mu^2}\ _{2}F_1(\frac{1}{2},\frac{D-3}{2};\frac{3}{2};-\frac{r_P^2-\mu^2}{\mu^2})\Big\}\nonumber\\&\simeq&\frac{\gamma}{4-D}\big(r_E^{4-D}+r_P^{4-D}\big),
%\end{eqnarray}

\noindent where $r_E~,~r_P\gg\mu$, and also the first two terms of expansion

\begin{eqnarray}\label{h92}
&& _{2}F_1(\frac{1}{2},\frac{D-3}{2};\frac{3}{2};-\frac{r^2-\mu^2}{\mu^2})\simeq\\ &&\frac{1}{4-D}(\frac{r}{\mu})^{3-D}-\frac{(D-3)}{2(2-D)}(\frac{r}{\mu})^{1-D}+\mathcal{O}\big((\frac{\mu}{r})^{1+D}\big),\nonumber
\end{eqnarray}

\noindent are used. It should also be noted that a one-way path has been assumed and, accordingly, for a round trip ($E\rightarrow P\rightarrow E$), the derived Shapiro time delay formulas must be duplicated.

In the Solar system tests and for the Cassini mission, while $M(\equiv M_\odot=1.9884\times10^{30}\textmd{Kg})$ denotes the Solar mass, $r_P(\sim8.43\textmd{AU})$ and $r_E(\sim1\textmd{AU})$ represent the spacecraft and Earth distances from the Sun, respectively. Additionally, we have $\mu\sim 1.6 R_\odot$, where $R_\odot(\sim0.005\textmd{AU})$ is the Solar radius \cite{Bertotti:2003rm, Moffat:2006gz, Kopeikin:2006yu, Will:2014kxa}. Moreover, $\frac{d\mu}{dt}$ is close to the Earth's velocity ($v_E=30~\textmd{km/s}$)  \cite{Bertotti:2003rm}. Observations also imply a peak ($6\times10^{-10}$) for the time changes of the Shapiro time delay ($\frac{d\delta T_{F}}{dt}$ and $\frac{d\delta T_{Sch}}{dt}$ in the fractional and non-fractional cases, respectively) \cite{Bertotti:2003rm}. Hence, for a round trip, Eq.~(\ref{h91}) generates

\begin{eqnarray}\label{h911}
\!\!6\times10^{-10}\simeq\frac{2\gamma\mu(D-3)}{c(D-2)}\big[(\frac{1}{r_E})^{D-2}+(\frac{1}{r_P})^{D-2}\big]v_E,
\end{eqnarray}

\noindent that finally { generates $D\simeq3.8994 \pm 10^{-10}$} in the unit of $G=6.6743\times10^{-11}~\textmd{m}^3\textmd{kg}^{-1}\textmd{s}^{-2}$ and $c=299792458~\textmd{m/s}$ where the extra $c$ appearing in the denominator is the result of employing the SI units ($\delta T_{F}\rightarrow c\delta T_{F}$). Consequently, more accurate measurements seem necessary to verify and study the fractional structure of the Sun.

%!!! The weak field limit of the ratio $\frac{\delta T_F}{\delta T_{Sch}}$ has been plotted as a function of $D$ in Fig.~\ref{fsch} for $G=1$. Consequently, $\frac{\delta T_F}{\delta T_{Sch}}$ versus $D$ is comparable with the Cassini spacecraft limitation ($\mathcal{O}(10^{-9})\textmd{sec}$) \cite{Moffat:2006gz, Kopeikin:2006yu}, a point signaling the power of the Shapiro test in verifying such a geometry. Correspondingly, the Shapiro test of the Solar system is fitted very well with $\delta T_{Sch}$, which indicates the non-fractional structure of the system. Therefore, time delays caused by strong gravitational fields, such as those of black holes, are required to use the Shapiro test in verifying the fractional structures.
%$D\simeq3.8759$
%Taking into account the expression~(\ref{h92}) up to the second term, then Eq.~%(\ref{h91}) is modified as

%\begin{eqnarray}\label{h93}
%\delta T_F&\simeq&\frac{\gamma}{4-D}\big(r_E^{4-D}+r_P^{4-D}\big)\\&+&\frac{\gamma\mu^2(D-3)}{2(D-2)}(r_E^{2-D}+r_P^{2-D}),\nonumber
%\end{eqnarray}

%\begin{figure}[H]
%\centering
%\includegraphics[width=0.8\linewidth]{FSCH.pdf}
%\caption{$\frac{\delta T_F}{\delta T_{Sch}}$ versus $D$ for $G=c=1$, $r_E\gg\mu$, and $r_P\gg\mu$.}\label{fsch}
%\end{figure}
%%%%%%%%%%%%%%%%%%%%%%%%%%%%%%%%%%%%%%%%%%%%%%%%%%%%%%%%%\textmd{\mu~sec}
\section{Bending of light and the orbital precession}

In addition to the time delay, the bending of null geodesics has other effects as well. It helps the central object $O$ to cast its shadow and bend the light rays. Moreover, the time-like geodesics (the planetary orbits) experience a precession. Here, we are going to study these effects for metric~(\ref{h2}).

\subsection{The Shadow}

To see an object, a photon emitted from it should be seen by the observer. If a secondary object absorbs the photon, the observer will not recognize the first object, and, in fact, the second object will cast its shadow on the background as well as in place of the first object. Correspondingly, a black hole has its own shadow whose size depends on the geometry and mass of the black hole and is calculable by finding the photon orbit ($r_{ph}$). For values larger (smaller) than $r_{ph}$, the corresponding photons will ultimately travel towards infinity (the black hole).

To go a step further, consider metric~(\ref{h2}) which has the two Killing vectors $\xi^\mu_1=\delta^\mu_t$ and $\xi^\mu_2=\delta^\mu_\phi$ corresponding to the two conserved quantities $E=-g_{tt}\dot{t}=f(r)\dot{t}$ and $L=g_{\phi\phi}\dot{\phi}=r^2\sin^2\theta~\dot{\phi}$, respectively, where "~$\dot{}$~" denotes derivative with respect to an affine parameter $\lambda$. Therefore, we have

\begin{equation}\label{sh1}
\begin{split}
&\dot{r}^2+V_{eff}=E^2,\\ &V_{eff}=\left(1-\frac{16\pi\tilde{G} M}{(D-2)r^{D-3}}\right)\frac{L^2}{r^2},
\end{split}
\end{equation}

\noindent for equatorial photons ($\theta=\frac{\pi}{2}$). The radius of the photon orbit (photon sphere) is obtained by solving, $\frac{dV_{eff}}{dr}=0$ leading to $r_{ph}=(\frac{16\pi\tilde{G}(D-1)M}{2(D-2)})^{\frac{1}{D-3}}$ combined with $r_{h}=(\frac{16\pi\tilde{G}M}{(D-2)})^{\frac{1}{D-3}}$ \cite{Jalalzadeh:2025uuv} to finally reach at $r_{ph}=(\frac{D-1}{2})^{\frac{1}{D-3}}r_h$. Moreover, since $3<D\leq4$, we have $\frac{d^2V_{eff}}{dr^2}\big|_{r=r_{ph}}<0$ that clarifies why the orbit $r=r_{ph}$ is unstable and forms a boundary between the photons absorbed by the black hole and those that run away to infinity. On the other hand, for this orbit, we have $\dot{r}=0$ combined with~(\ref{sh1}) to get the impact parameter $b(\equiv\frac{L}{E})$ \cite{Chandrasekhar:1985kt} corresponding to radius $r_{ph}$ as

\begin{eqnarray}\label{sh2}
b_{ph}=\frac{r_{ph}}{\sqrt{1-\frac{16\pi\tilde{G}M}{(D-2)r_{ph}^{D-3}}}}=\sqrt{\frac{D-1}{D-3}}~r_{ph}.
\end{eqnarray}

\noindent Clearly, a photon with angular momentum $\mathcal{L}$, energy $\mathcal{E}$, and thus the impact parameter $b=\frac{\mathcal{L}}{\mathcal{E}}$ will be absorbed by a black hole with the impact parameter $b_{ph}$ or will move away from it if $b<b_{ph}$ or $b_{ph}<b$, respectively. Loosely speaking, $b_{ph}$ determines the shadow size for a distant observer \cite{Chandrasekhar:1985kt} and the Schwarzschild results (where $b_{ph}=\sqrt{3}~r_{ph}$, $r_{ph}=3/2~r_h$, and $r_h=2GM$) are easily obtainable when $D=4$ \cite{Chandrasekhar:1985kt}.

\begin{figure}
    \centering
    \includegraphics[width=1\linewidth]{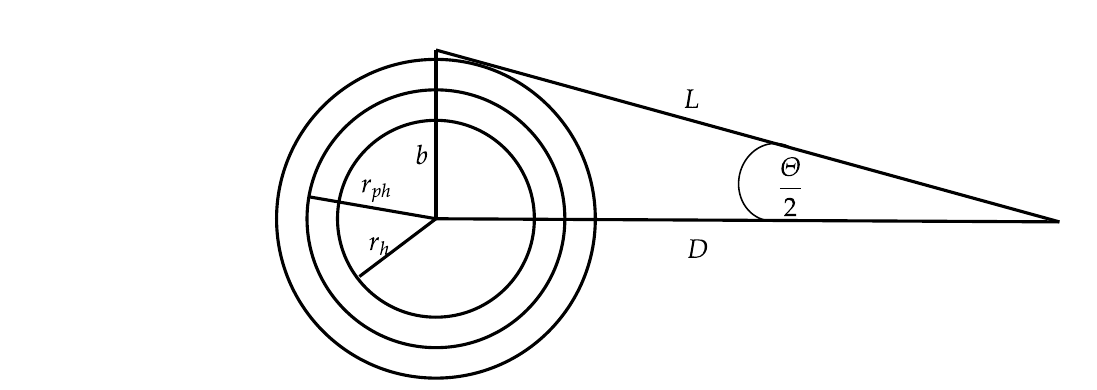}
    \caption{Approximate relationship between the impact parameter, $D$ (black hole distance), and $\Theta$.}
    \label{shad}
\end{figure}

Consider $\textmd{M}87$ with crescent diameter $\Theta=42\pm3~\mu$as located at distance $D\simeq16.8$Mpc \cite{EventHorizonTelescope:2019ggy}. Therefore, in full accordance with $\Theta\ll1$, we have $b\ll D$ that finally guides us to $\tan\Theta\sim\Theta$ (Fig.~\ref{shad}) and an estimation for the observational value of the impact parameter as

\begin{eqnarray}\label{sh201}
&&b_{ob}\approx\frac{\Theta}{2}D\simeq(21\times10^{-6}\times4.84814\times10^{-6})\\&&\times~(16.8\times10^{6}\times3.086\times10^{16})\approx5.28\times10^{13}\textmd{m}.\nonumber
\end{eqnarray}

\noindent By bearing the SI units in mind, we arrive at $\frac{b_{ob}}{\frac{GM}{c^2}}\approx5.558$ for $M=6.5\times10^9M_\odot$, and correspondingly, $GM/c^2\sim0.95\times10^{13}\textmd{m}$ \cite{EventHorizonTelescope:2019ggy}. Theoretical predictions based on the Schwarzschild metric ($D=4$) yield $\frac{b}{GM/c^2}=3\sqrt{3}\simeq5.196$, and to verify the potential of fractionality in filling this gap, we use Eq.~(\ref{sh2}) to get

\begin{eqnarray}\label{sh202}
&&\frac{b_{ph}}{GM/c^2}=\\&&\sqrt{\frac{D-1}{D-3}}\big[\frac{16\pi\mathcal{A}(D-1)}{2(D-2)}\big]^{\frac{1}{D-3}}(\frac{\frac{GM}{c^2}}{l_p})^{\frac{4-D}{D-3}},\nonumber
\end{eqnarray}

\noindent { yielding $D\simeq3.9994 \pm 7 \times 10^{-4}$} for $\textmd{M}87$ and $l_p=1.6162\times10^{-35}\textmd{m}$. Therefore, fractionality, its quality, and implications in black holes deserve further serious study and observations.

%$D\simeq3.9766$
%%%%%%%%%%%%%%%%%%%%%%%%%%%%%%%%%%%%%%%%%%%%%%%%%%%%%%%%%
\subsection{Deflection angle}

\begin{figure}
    \centering
    \includegraphics[width=0.9\linewidth]{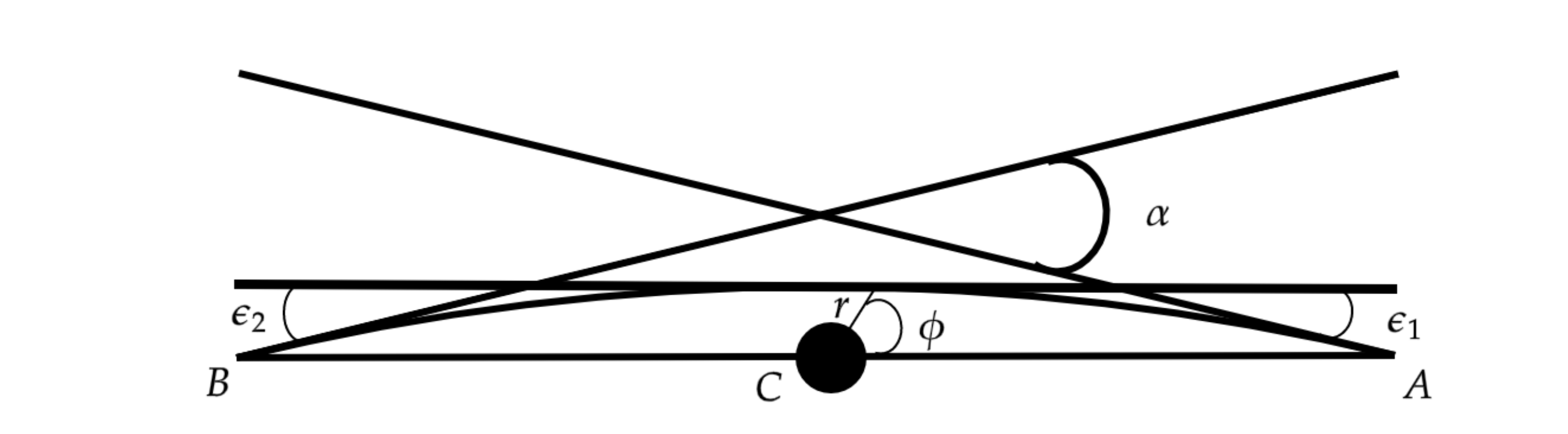}
    \caption{In the presence of an object, the null geodesics are bent, and correspondingly, the observer $B$ can see the object $A$ but in another position, i.e., a straight path extension with a deflection angle $\alpha$.}
    \label{GL1}
\end{figure}

On the $\theta=\frac{\pi}{2}$ surface, where $L=g_{\phi\phi}\dot{\phi}=r^2\dot{\phi}$, using the definition $u=\frac{1}{r}$ in rewriting Eq.~(\ref{sh1}), one gets

\begin{eqnarray}\label{d1}
(\frac{du}{d\phi})^2+u^2=\frac{1}{b^2}+\gamma u^{D-1}\Rightarrow\frac{d^2u}{d\phi^2}+u=3\eta u^{D-2},
\end{eqnarray}

\noindent whose solution (up to the first order of $\eta$) can be obtained by solving

\begin{eqnarray}\label{d2}
&&\frac{d^2u_0(\phi)}{d\phi^2}+u_0(\phi)=0,\nonumber\\ &&u_1^{\prime\prime}(\phi)+u_1(\phi)=3 u_0^{D-2}(\phi),
\end{eqnarray}

\noindent where $^\prime\equiv\frac{d}{d\phi}$ and the expansion $u(\phi)\simeq u_0(\phi) +\eta u_1(\phi) +\mathcal{O}(\eta^2)$ in which $\eta\equiv\frac{\gamma(D-1)}{6}$, has been employed. To solve the first equation, note that the closest distance of a photon from the object $C$ (it goes from the distant source $A$ to the distant observer $B$) happens at $\phi=\frac{\pi}{2}$ (Fig.~(\ref{GL1})). Indeed, it is the radius of the object $C$ observed by $B$ ($r(\frac{\pi}{2})=b=\frac{1}{u_0(\frac{\pi}{2})}$), $u^\prime_0(\frac{\pi}{2})=0$ and thus $u_0(\phi)=u_0\sin\phi=\frac{\sin\phi}{b}$. For $D=4$ and up to the first order of $M$, since $\sin^2\phi=\frac{1-\cos2\phi}{2}$, one reaches at

\begin{eqnarray}\label{d3}
u(\phi)=\frac{\sin\phi}{b}+\frac{GM}{2b^2}(3+\cos2\phi).
\end{eqnarray}

\noindent It is also easy to verify that the minimum remains at $\phi=\frac{\pi}{2}$, where $u^{\prime}(\phi=\frac{\pi}{2})=0$. To find deflection angle ($\alpha\equiv\Delta\phi-\pi$), two conditions including $i$) $u(\phi\equiv\epsilon_1\simeq0)=0$ leading to $\sin\phi\simeq\epsilon_1$ and $ii$) $u(\phi\simeq\pi+\epsilon_2)=0$ for which, $\sin\phi\simeq-\epsilon_2$ should be employed that equip us with

\begin{eqnarray}\label{d4}
\bigg\{^{\epsilon_1=-\frac{2GM}{b}}_{\epsilon_2=\frac{2GM}{b}}\rightarrow\Delta\phi=\pi+\epsilon_2-\epsilon_1=\pi+\frac{4GM}{b},
\end{eqnarray}

\noindent and thus $\alpha=\frac{4GM}{b}$. For other values of $D$, it is a matter of calculation to show

\begin{eqnarray}\label{d5}
u(\phi)&=&\frac{\sin\phi}{b}+\frac{3\eta}{b^{D-2}}\bigg[\frac{\sin^D\phi}{D-1}\nonumber\\&+&\cos\phi\int_{\phi}^\frac{\pi}{2}\sin^{D-1}s~ds\bigg],
\end{eqnarray}

\noindent that recovers~(\ref{d3}) for $D=4$. In looking for the deflection angle corresponding to the general value of $D$, by following the above recipe and only keeping the terms up to the first order $\epsilon_1$ and $\epsilon_2$, one reaches

\begin{eqnarray}\label{d6}
\alpha_D=\frac{\gamma(D-1)B(\frac{D}{2},\frac{1}{2})}{2b^{D-3}},
\end{eqnarray}

\noindent recovering $\alpha=\frac{4GM}{b}$ as a desired limit when $D=4$ for which $B(2,\frac{1}{2})=\frac{4}{3}$ and $\gamma=2GM$.

While Newtonian calculations and even the usage of special relativity produce one-half of the general relativity (GR) result, in the SI unit, GR predicts $\alpha=\frac{4GM}{bc^2}\approx1.75$ arc-sec and the Eddington's team reports $\alpha\approx1.61\pm0.30$ arc-sec \cite{def1, def2, def3, def4}. Therefore, having in hand the radius of the Sun as the impact parameter and accepting the observational report $\alpha\approx1.61\pm0.30$ arc-sec \cite{def1, def2, def3, def4}, Eq.~(\ref{d6}) leads { to $D\simeq3.8780 \pm 9 \times 10^{-3}$.}

So far, only the weak field limit of Eq.~(\ref{sh1}) has been studied, which led us to~(\ref{d6}). Indeed, for a non-weak field limit and on an equatorial plane, a static spherically symmetric metric like~(\ref{h2}), produces the deflection angle \cite{Weinberg:1972kfs}

\begin{eqnarray}\label{d7}
\alpha(D)=2\int_{r_0}^\infty\frac{dr}{r^2\sqrt{\frac{1}{b^2}-\frac{f(r)}{r^2}}}-\pi,
\end{eqnarray}

\noindent obtained by inserting $\frac{dr}{d\phi}=\frac{r^2}{L}\dot{r}$ into Eq.~(\ref{sh1}). Here, $r_0$ denotes the closest distance of the photon from the central object, and Eq.~(\ref{d6}) (weak field limit) is also recovered by expanding~(\ref{d7}) up to the first order $\eta$. For more intuition, it is easily verifiable for $D=4$.

%%%%%%%%%%%%%%%%%%%%%%%%%%%%%%%%%%%%%%%%%%%%%%
\subsection{Precession of orbits}

In Newtonian gravity, a planet with mass $m$ and energy $E$ follows an elliptical orbit for which $\mathbb{E}\equiv\frac{E}{m}=\frac{1}{2}m(\frac{dr}{dt})^2+\frac{L^2}{2mr^2}-\frac{GMm}{r}$ and $L=\sqrt{GMa(1-e^2)}$ denote angular momentum and energy per mass, respectively, and \cite{fowls}

\begin{eqnarray}\label{n1}
&&r(\phi)=\frac{\frac{L^2}{GM}}{\big[1+e\cos(\phi)\big]}=\frac{a(1-e^2)}{\big[1+e\cos(\phi)\big]},%\\ %&& %\big<\frac{1}{r(\phi)}\big>=\frac{1}%{2\pi}\int_0^{2\pi}\frac{\big[1+e\cos(\phi)\big]}{a(1-e^2)}d\phi=\frac{1}{a(1-%e^2)},\nonumber
\end{eqnarray}

\noindent is the orbit equation. Here, $e$ and $a$ represent the eccentricity and half of the larger diameter of the elliptical orbit, respectively. By combining the law of conservation of energy and the amount of energy at the apogee and perigee points (for which $\frac{dr}{dt}=0$), as well as the geometry of elliptical orbits (it allows us to establish relationships between the radii of the apogee and perigee points and $e$), one can obtain $L=\sqrt{GMa(1-e^2)}$ in agreement with Eq.~(\ref{n1}) \cite{fowls}. Finally, it should be noted that since $r(\phi)=r(\phi+2\pi)$, there is no orbit precession in Newtonian gravity where space is flat.

Metric~(\ref{h2}) enables one to reach

\begin{eqnarray}\label{p1}
&&(\frac{dr}{d\tau})^2=E^2-V_{eff}(r),\\ && V_{eff}(r)=f(r)(1+\frac{L^2}{r^2}),\nonumber
\end{eqnarray}

\noindent where $L=r^2\frac{d\phi}{d\tau}$ and $E=-f(r)\frac{dt}{d\tau}$ are constants of motion. Now, by defining $\mathcal{R}\equiv\frac{1}{r}$ and using $\frac{dr}{d\tau}=\frac{dr}{d\phi}\frac{d\phi}{d\tau}$, we easily obtain

\begin{eqnarray}\label{p2}
\frac{d^2\mathcal{R}}{d\phi^2}+\mathcal{R}=\frac{3(D-2)\eta}{(D-1)L^2}\mathcal{R}^{D-4}+3\eta\mathcal{R}^{D-2}.
\end{eqnarray}

\noindent The Schwarzschild result and Newtonian regime appear for $D=4$ and when in addition to $D=4$, the term $3\eta\mathcal{R}^{D-2}$ is also ignored, respectively. The general solution is equal to

\begin{eqnarray}\label{p3}
&&\int_1^{\mathcal{R}(\phi)}\frac{d\xi^2}{\sqrt{\frac{6(D-2)\eta\xi^{D-3}}{(D-1)(D-3)L^2}+\frac{6\eta\xi^{D-1}}{D-1}+c_1-\xi^2}}\nonumber\\&&=(c_2+\phi)^2,
\end{eqnarray}

\noindent where $c_i$ are the constants of integration. For $D=4$ and up to the lowest order of $L^{-2}$, Eq.~(\ref{p2}) clearly produces the Schwarzschild expectation where \cite{Weinberg:1972kfs}

\begin{eqnarray}\label{p4}
&&\mathcal{R}_{D=4}(\phi)\simeq \frac{GM}{L^2}\big[1+e\cos((1-\delta_{D=4})\phi)\big],\nonumber\\ &&\delta_{D=4}=3\frac{G^2M^2}{L^2},~\Delta\phi_{D=4}=\frac{6\pi GM}{a(1-e^2)}.
\end{eqnarray}
%This is available when $M^2 \ll L^2$.
\noindent Since Eq.~(\ref{n1}) should be recovered when $\delta_{D=4}$ is ignorable, the first approximation becomes $L^2\simeq GMa(1-e^2)$. Indeed, by solving $\frac{dV_{eff}(r)}{dr}\big|_{D=4}=0$ and using $r=a(1-e^2)$, one reaches at $L=\sqrt{\frac{GMr^2}{r-3M}}=\sqrt{GMr\big(1+\mathcal{O}(\frac{M}{r})\big)}\simeq\sqrt{GMa(1-e^2)}$ \cite{Weinberg:1972kfs}.

Obviously, the solution of Eq.~(\ref{p2}) must be reduced to Eq.~(\ref{p4}) for $D=4$. Now, at the weak field limit (like the Solar system), to approximately solve Eq.~(\ref{p2}) up to the lowest order of $L^{-2}$, it is a matter of calculation to reach

\begin{eqnarray}\label{p5}
&&\mathcal{R}_D(\phi)\simeq\frac{GM}{L^2}\big[1+e\cos((1-\delta_D)\phi)\big],\\ &&\delta_D=\frac{3\eta(D-2)}{2}(\frac{GM}{L^2})^{D-3},\nonumber
\end{eqnarray}

\noindent as the counterpart of $\mathcal{R}_{D=4}(\phi)$ for $D\neq4$. Moreover, we have $\eta=GM$ and thus $\delta_D\rightarrow3\frac{G^2M^2}{L^2}$ when $D=4$. For the angular momentum, one can easily get

\begin{eqnarray}\label{p6}
\frac{dV_{eff}(r)}{dr}=0\rightarrow L^2=\frac{3(D-3)\eta r^2}{(D-1)\big(r^{D-3}-3\eta\big)},
\end{eqnarray}

\noindent combined with $r=a(1-e^2)$ \cite{Weinberg:1972kfs} to find

\begin{eqnarray}\label{p61}
L^2\simeq\frac{3(D-3)\eta}{(D-1)}\big(a(1-e^2)\big)^{5-D},
\end{eqnarray}

\noindent up to the first order of $\eta$. Consequently, using $\gamma(D=4)\equiv\gamma_{4}=2GM$, for the orbit equation and precession, we have

\begin{eqnarray}\label{p7}
&&\mathcal{R}_D(\phi)\simeq\frac{\gamma_4\big[1+e\cos((1-\delta_D)\phi)\big]}{\gamma(D-3)\big(a(1-e^2)\big)^{5-D}},\nonumber\\
&&\Delta\phi_D\approx~2\pi\delta_D=\frac{\pi\gamma(D-2)}{\gamma_{4}}\times\\&&\big[\frac{\gamma_{4}}{\gamma(D-3)\big(a(1-e^2)\big)^{5-D}}\big]^{D-2}\bigg\{\frac{\gamma(D-3)(D-1)}{2\big(a(1-e^2)\big)^{D-5}}\bigg\},\nonumber
\end{eqnarray}

\noindent respectively, that clearly recover Eq.~(\ref{p4}) for $D=4$. The Newtonian result is also automatically achieved when $D=4$ and $\delta_D$ becomes ignorable.

In the case of Mercury, where $a(1-e^2)=55.46\times10^{9}$ m, while observations report $\Delta\phi_O=43.11\pm0.45$ arc seconds per century (as/cy) for the perihelion advance, by employing Eq.~(\ref{p4}), the best theoretical estimations are $\Delta\phi_{D=4}=43.03$ as/cy \cite{Cel}, $\Delta\phi_{D=4}=42.98$ as/cy \cite{Nobili}, and $\Delta\phi_{D=4}=43.2$ as/cy \cite{Berche:2024cwe}. Now, by considering the natural units system and using $M=M_\odot$, Eq.~(\ref{p7}) would be aligned with observations provided { that $D\simeq3.9949 \pm 2 \times 10^{-2}$.} Despite the approximations used in obtaining the value of $D$ corresponding to the Sun by using the deflection angle ($D\simeq3.8780$), Eqs.~(\ref{h911}), and~(\ref{p7}), the results are close to each other. Consequently, investigating the existence, quality, and amount of fractionality in the Sun is a topic that deserves investigation.

%%%%%%%%%%%%%%%%%%%%%%%%%%%%%%%%%%%%%%%%%%%%%%%%%%%%%%%%%%%%%%%%%%%%%%%%
\section{Bayesian MCMC Analysis}

{ Motivated by the obtained values of $D$ and their difference from $4$ (the Schwarzschild limit), here, we are going to employ the MCMC method to constrain the fractional parameter $D$ and to fit its theoretical predictions to observational data.

Assuming Gaussian distributed observational uncertainties, the likelihood function admits the following relation with the effective $\chi^2$ function
\begin{equation}
\mathcal{L} \propto \exp\!\left(-\frac{1}{2}\chi^2\right),
\end{equation}

\subsection{Solar System constraints}
As we investigated in previous sections, our first analysis is based on three independent tests in the solar system, including the Shapiro time delay, the deflection angle, and the precession of orbits. For the deflection angle, and the precession of orbits, the $\chi^2$ function is defined as
\begin{equation}
\chi^2_i(D) =
\frac{\left[X_{\mathrm{th}}(D) - X_{\mathrm{obs}}\right]^2}
{\sigma_i^2},
\end{equation}

\noindent where $X_{\mathrm{th}}(D)$ is the theoretical prediction of the model ($\alpha$ from Eq.~\ref{d6} for the deflection angle and $\Delta\phi$ from Eq.~\ref{p7} for the precession of orbits), $X_{\mathrm{obs}}$ is the observational value ($1.61$ for deflection angle and $43.11$ for the precession of orbits), and $\sigma_i$ denotes the observational uncertainty ($0.30$ for deflection angle and $0.45$ for the precession of orbits).

The Shapiro time delay measurement has an extremely small uncertainty of order $\sigma_{\mathrm{obs}} \sim 10^{-15}$~\cite{Kliore:2004}. Therefore, it may artificially overconstrain the parameter $D$.
To address this issue, we introduce a Bayesian parameter $\lambda$ that rescales the effective uncertainty as $\sigma_{\mathrm{eff}} = \lambda \sigma_{\mathrm{obs}}$. The parameter of $\lambda$ is considered as an additional free parameter and is sampled with $D$ in MCMC analysis. The $\chi^2$ function for the Shapiro time delay is

\begin{equation}
\chi^2_{\mathrm{Shapiro}}(D,\lambda) =
\frac{\left[\delta T_{\mathrm{th}}(D) - \delta T_{\mathrm{obs}}\right]^2}
{(\lambda \sigma_{\mathrm{obs}})^2}
+ 2 \ln \lambda,
\end{equation}

\noindent where the logarithmic term arises from the normalization of the Gaussian likelihood and acts as a Bayesian penalty against arbitrarily large values of $\lambda$~\cite{Hobson:2002}. We use \texttt{emcee} package~\cite{Foreman-Mackey:2013} and perform four separate MCMC analyses including deflection angle (1D sampling in $D$), the precession of orbits (1D sampling in $D$), shapiro time delay (2D sampling in $(D,\lambda)$), and a combined analysis that includes all three observables (2D sampling in $(D,\lambda)$). For all analyses, we use $3 < D \leq 4$, and $0.1 < \lambda < 100.$ ranges.

Figure~\ref{orbital precession} shows the one-dimensional marginalized posterior distribution of the fractional dimension $D$ inferred from the perihelion precession data. The blue histogram represents the posterior samples obtained from the MCMC analysis after converging. The solid black vertical line denotes the median value of the posterior $D=3.83 \pm 0.07$, the two dotted black lines indicate the corresponding $1\sigma$ credible interval. The red dashed vertical line marks the GR prediction $D=4$. Although the posterior is broader compared to the deflection constraint, reflecting the weaker constraining power of the perihelion precession data, the inferred value of $D$ remains statistically consistent with GR.

\begin{figure}
    \centering
    \includegraphics[width=0.9\linewidth]{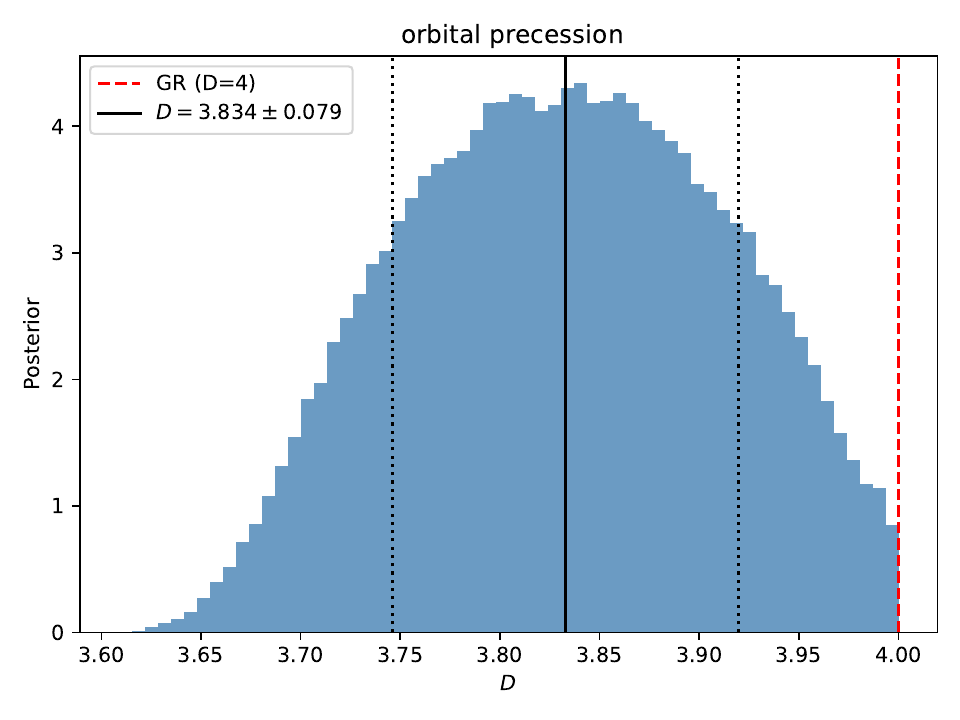}
    \caption{One-dimensional marginalized posterior distribution of the fractional dimension $D$ obtained from perihelion precession data.}
    \label{orbital precession}
\end{figure}

Figure~\ref{Deflection angle} presents a one-dimensional marginalized posterior distribution of the fractional dimension $D$ inferred from the deflection angle data. The blue histogram shows the posterior samples obtained from the MCMC analysis after burn-in and convergence. The solid black vertical line indicates the median value of the posterior distribution ($D=3.995 \pm 0.003$), the two dotted black lines mark the corresponding $1\sigma$ credible interval. The red dashed vertical line denotes the General Relativity prediction $D=4$. In contrast to the perihelion precession case, the posterior distribution is sharply peaked, reflecting the strong constraining power of deflection angle measurements on the fractional dimension. The inferred value of $D$ lies extremely close to the General Relativity limit $D=4$.

\begin{figure}
    \centering
    \includegraphics[width=0.9\linewidth]{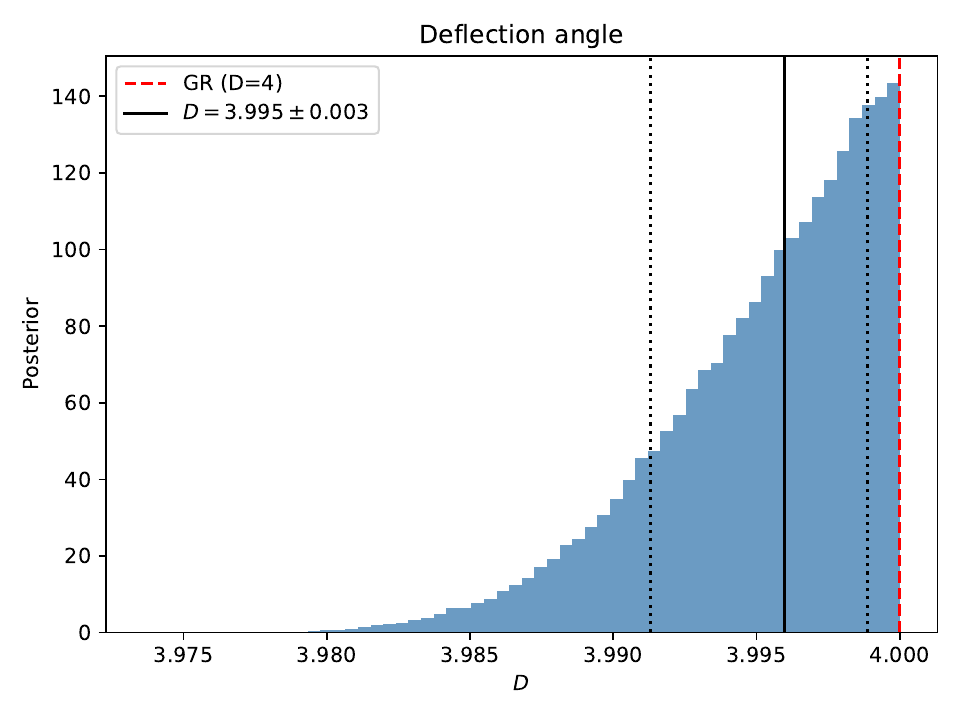}
    \caption{One-dimensional marginalized posterior distribution of the fractional dimension $D$ obtained from deflection angle data.}
    \label{Deflection angle}
\end{figure}

%%%%%%%%%%%%%%%%%%%%%%%%%%%%%%%%%%%%%%%%%%%%%%%%%%%
%\onecolumn
%\begin{multicols}{1}
\begin{table*}[ht]
\centering
\caption{A comparative table including a summary on Schwarzschild-Tangherlini (SchwT) and fractional Schwarzschild-Tangherlini (FSchwT) black holes.}
\label{table:1}
\begin{tabular}{|c|c|c|}
\hline
$\dfrac{\textmd{Black~hole}~\Rightarrow}{\textmd{Tests}~\Downarrow}$ & SchwT ($D=4$) & FSchT ($3<D\leq4$) \\ \hline

Shapiro time &
\makecell{$2GM\ln\left\{(\sqrt{r_E^2-\mu^2}+r_E)(\sqrt{r_P^2-\mu^2}+r_P)/\mu^2\right\}$} &
\makecell{$\dfrac{\gamma}{\mu^{D-3}}\Big\{\sqrt{r_E^2-\mu^2}\, {}_{2}F_1\!\left(\dfrac{1}{2},\dfrac{D-3}{2};\dfrac{3}{2};-\dfrac{r_E^2-\mu^2}{\mu^2}\right)$\\
$+\sqrt{r_P^2-\mu^2}\, {}_{2}F_1\!\left(\dfrac{1}{2},\dfrac{D-3}{2};\dfrac{3}{2};-\dfrac{r_P^2-\mu^2}{\mu^2}\right)\Big\}$} \\ \hline

Sagnac time delay &
\makecell{$A\omega_0\left(1-\dfrac{2M}{R}-R^2\omega_0^2\right)^{-1/2}$} &
\makecell{$A\omega_0\left(1-\dfrac{\gamma}{R^{D-3}}-R^2\omega_0^{\,2}\right)^{-1/2}$} \\ \hline

\makecell{Bending of light\\(weak field limit)} &
$\dfrac{4GM}{b}$ &
$\dfrac{3B\left(\dfrac{D}{2},\dfrac{1}{2}\right)\eta}{b^{D-3}}$ \\ \hline

\makecell{Photon sphere $\&$\\ the shadow size} &
\makecell{$b_{ph}=3\sqrt{3}GM,\quad r_{ph}=3GM,$\\ $r_h=2GM$} &
\makecell{$b_{ph}=\sqrt{\dfrac{D-1}{D-3}}\,r_{ph},\quad
r_{ph}=\left(\dfrac{D-1}{2}\right)^{\frac{1}{D-3}}r_h,$\\
$r_{h}=\left(\dfrac{16\pi\tilde{G}M}{D-2}\right)^{\frac{1}{D-3}}$} \\ \hline

\makecell{Precession of orbit\\(weak field limit)} &
$\dfrac{6\pi GM}{a(1-e^2)}$ &
\makecell{$\dfrac{\pi\gamma(D-2)}{\gamma_{4}}
\left[\dfrac{\gamma_{4}}{\gamma(D-3)\big(a(1-e^2)\big)^{5-D}}\right]^{D-2}$\\
$\times\left\{\dfrac{\gamma(D-3)(D-1)}{2\big(a(1-e^2)\big)^{D-5}}\right\}$} \\ \hline

\end{tabular}
\end{table*}

%%%%%%%%%%%%%%%%%%%%%%%%%%%%%%%%%%%%%%%%%%%%%%%%%%%%%%%%%

Figure~\ref{shapiro corner} shows the posterior distribution of the fractional dimension $D$ and the parameter $\lambda$ inferred from the Shapiro time delay data. The upper left panel shows the posterior distribution of $D$. The median of the marginalized posterior's value is $D = 3.700^{+0.204}_{-0.122}$. The posterior distribution for the fractional dimension is broad and asymmetric. The value $D=4$ is located in the tail of the distribution. The Shapiro time delay data alone cannot constrain the parameter $D$ with high accuracy. The lower right panel shows the posterior distribution of $\lambda$. The value of $\lambda$ is equal to $57.71^{+27.66}_{27.94}$ and thus, $\sigma_{\mathrm{eff}} \approx 60 \times 10^{-15}$. The lower left panel shows the correlation between $D$ and $\lambda$ where it is seen that large values of $\lambda$ are required to have $D$ near $4$. For a small value of $\lambda$ parameter, $\lambda$ sticks to the edge of the prior. Therefore, without $\lambda$, the median of the marginalized posterior of $D$ sticks to the prior.

\begin{figure}
    \centering
    \includegraphics[width=0.9\linewidth]{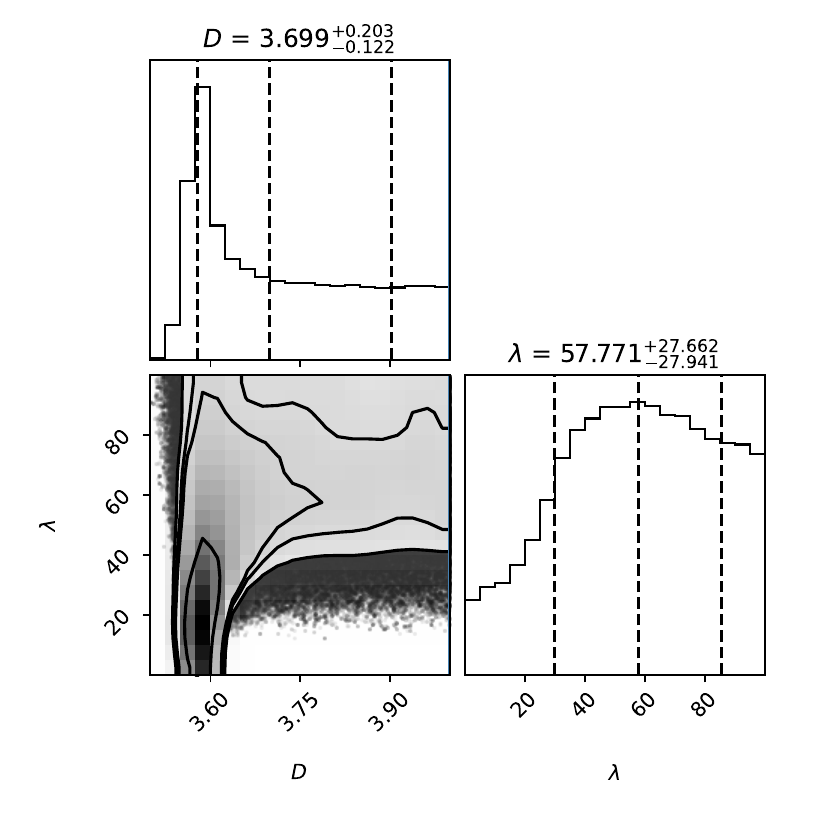}
    \caption{Corner plot showing the posterior distribution of the fractional dimension $D$ and the parameter $\lambda$ obtained from Shapiro time delay data.}
    \label{shapiro corner}
\end{figure}
\begin{figure}[!ht]
    \centering
    \includegraphics[width=0.9\linewidth]{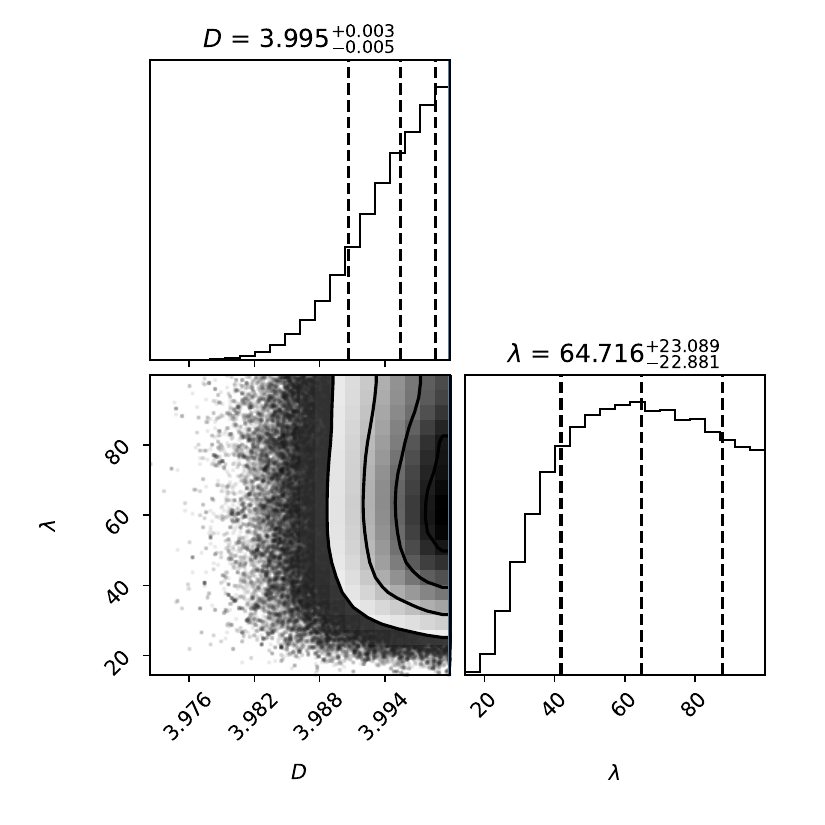}
    \caption{Corner plot showing the posterior distribution of the fractional dimension $D$ and the parameter $\lambda$ obtained from combined data.}
    \label{combined corner}
\end{figure}

Figure~\ref{combined corner} shows the posterior distribution of the fractional dimension $D$ and the parameter $\lambda$ inferred from the combined data. The upper left panel shows the posterior distribution of $D$. The middle dashed line shows the median of the posterior distribution's value of $D = 3.99^{+0.003}_{-0.005}$. Right and left dashed lines show the distribution of $1\sigma$. The distribution is very sharp and asymmetric. Therefore, Solar System observational data are consistent with $D = 3.99^{+0.003}_{-0.005}$. The lower right panel shows the posterior distribution of $\lambda$. The median of the posterior distribution's value of $\lambda$ is equal to $64.71^{+23.08}_{22.88}$. The lower left panel shows the correlation between the parameters $D$ and $\lambda$. As $D$ moves away from $4$, the model can explain the Shapiro time delay data by increasing the value of $\lambda$.

\subsection{Black hole shadow constraint}

In the black hole shadow analysis, the predicted shadow size depends
on the spacetime dimension $D$ through a power law factor of the form
$
\left(\frac{GM}{c^2 l_p}\right)^{\frac{4-D}{D-3}} .
$
As a result, even a tiny deviation from the general relativistic
value ($D=4$) for $D$ leads to a dramatic mismatch between the model prediction
and the observed shadow size, causing the corresponding $\chi^2$
to grow effectively without bounds. Therefore, the black hole shadow data do not provide a meaningful statistical constraint on $D,$ meaning that the studied metric cannot provide a proper model for $\textmd{M}87$, at least within the current data range.}

%%%%%%%%%%%%%%%%%%%%%%%%%%%%%%%%%%%%%%%%%%%%%%%%%%%%%%

%%%%%%%%%%%%%%%%%%%%%%%%%%%%%%%%%%%%%%%%%%%%%%%%%%%%%%%%%%%%%%%%%%%%%%%%

\section{Conclusions}

Having in hand a recently introduced line element including a fractional Schwarzschild-Tangherlini black hole with a fractal event horizon, we have tried to study the possibility of the existence of fractionality in the Sun and the $\textmd{M}87$. While the Shapiro time delay, deflection angle, and the precession of the Mercury orbit pave our way to study the Sun, $\textmd{M}87$ has been surveyed using its shadow. Differences with the ordinary Schwarzschild spacetime have been summarised in Table~$\textmd {I}$. { Overall, the MCMC analysis reveals the potential of the fractional Schwarzschild-Tangherlini line element in meeting the Solar tests. Moreover, the MCMC analysis of $\textmd{M}87$ can be considered as a motivation to look for and study other fractional black hole solutions in the hope that they can meet observational expectations. Correspondingly, the potential of fractionality in modeling various objects deserves further study.}

%%%%%%%%%%%%%%%%%%%%%%%%%%%%%%%%%%%%%%%%%
%%%%%%%%%%%%%%%%%%%%%%%%%%%%%%%

\section*{DATA AVAILABILITY}

The data used in this work are available publicly.

%%%%%%%%%%%%%%%%%%%%%%%%%%%%%%%%%%%%%%%%%%%%%%%%%%%%%%%%%%%%%%%%%%%
\section*{Declaration of competing interest}
The authors declare that they have no known competing financial
interests or personal relationships that could have appeared to
influence the work reported in this paper.

%%%%%%%%%%%%%%%%%%%%%%%%%%%%%%%%%%%%%%%%%%%%%%%%%%%%%%%%%%%%%%%%%%%%%%%%%
%\section*{Acknowledgements}
%The author would like to thank the anonymous reviewer for his/her insightful suggestions and careful manuscript reading.

%%%%%%%%%%%%%%%%%%%%%%%%%%%%%%%%%%%%%%%%%%%%%%%%%%%%%%%%%%%%%%%%%%%

%\section{Appendix title 2}

%\bibliographystyle{elsarticle-num} 
%\bibliography{example}



%% For example:

%% \bibitem[Aladro et al.(2015)]{Aladro15} Aladro, R., Martín, S., Riquelme, D., et al. 2015, \aas, 579, A101

%%\end{thebibliography}

\end{document}